\documentclass[12pt,a4paper]{article}
\usepackage[utf8]{inputenc} 
\usepackage[T2A]{fontenc} 
\usepackage{amsmath}
\usepackage{authblk}
\usepackage{mathtools}
\usepackage{ dsfont }
\usepackage{ wasysym }
\usepackage{braket}
\usepackage{amsfonts}
\usepackage{amssymb}
\usepackage{ upgreek }
\usepackage[unicode, pdftex]{hyperref}

\usepackage{xcolor}
\usepackage{hyperref}
\usepackage{graphicx}
\usepackage{pst-solides3d}
\usepackage{float}
\usepackage[left=2.7cm,right=2.7cm,
    top=1.5cm,bottom=1.5cm,bindingoffset=0cm]{geometry}

\author[1,2]{Alexander Belavin}
\author[1,3]{Boris Eremin}
\affil[1]{\textit{Landau Institute for Theoretical Physics, \protect 142432 Chernogolovka, Russia} } 
\affil[2]{\textit{Kharkevich Institute for Information Transmission Problems,\protect  127994 Moscow, Russia}}
\affil[3]{\textit{Moscow Institute of Physics and Technology,\protect 141700 Dolgoprudny, Russia.}}
\title{Partition function of $\mathcal{N}=(2,2)$ supersymmetric sigma models and Special geometry 
for the two-moduli  non-Fermat  Calabi-Yau manifold}
\date{\today} 
\begin{document}

\maketitle

\begin{center}
{\bf Abstract}
\end{center}
\par We study the new case of the application of the JKLMR conjecture on the connection between the exact partition functions of $\mathcal{N}=(2,2)$ supersymmetric gauged linear sigma models (GLSM) on $S^2$ and special Kähler geometry on the moduli spaces of Calabi-Yau manifold $Y$. The last ones arise as  manifolds of the supersymmetric vacua  of the GLSM. We establish this correspondence using the Mirror symmetry in Batyrev's approach. Namely, starting from  the two-moduli non-Fermat Calabi-Yau manifold $X$ we construct the dual GLSM with the supersymmetric vacua $Y$, which is the mirror for $X$. Knowing the special geometry on the complex moduli space of $X$ we verify the mirror version  of the JKLMR conjecture  by explicit computation.
\vspace{1.0 cm}


\section{Introduction}

\qquad Superstring theory is considered as a possible approach for unifying the Standard model and Quantum gravity. To obtain $4d$ theory with Spacetime supersymmetry (which is needed for the phenomenological reasons) we have to compactify 6 of 10 dimensions of Superstring theory on Calabi-Yau manifolds $X$ \cite{Super1, Green}. The resulting Lagrangian of the low-energy effective theory is defined by so called Special  geometry which appears on the moduli space of CY manifold $X$ \cite{Strominger, ModuliSpace, CandelasOssa}. Indeed the moduli space of Calabi-Yau manifold $X$ is a product of two factors: Moduli space $M_K(X)$ of the Kähler structure deformations and Moduli space of the complex structure deformations $M_C(X)$. Therefore for finding the Effective low energy theory we have to compute the Special Kähler geometry on the both Moduli spaces of CY manifolds.
\par A new approach for computing the Special geometry on the moduli space of complex structures 
$M_K(X)$  has been  introduced in \cite{Approach} some time ago. 
This method is based on the isomorphism between the middle cohomologies on CY and Chiral ring defined by the polynomial $W_X$ whose zero locus is the CY hypersurface $X$ in the weighted projective space.
\par On the other hand the conjecture for the explicit expression for the Kähler potential  
for the moduli space  of the Kähler structure deformations $M_K(Y)$
 was suggested and checked recently \cite{Jockers}. 
This conjecture (JKLMR conjecture) is the equality
\begin{equation}
\label{eq:conj}
  e^{-K_K^Y}=Z_Y,  
\end{equation}
 where $K_K^Y$ is the Kähler potential of the special geometry on the Kähler moduli spaces of Calabi-Yau manifold $Y$ defined as a hypersurface in the toric variety. The $Z_Y$ is the partition function of some  $\mathcal{N}=(2,2)$ gauge linear sigma model (GLSM) \cite{Witten} on $S^2$. Partition function $Z_Y$ was computed exactly by the Supersymmetric localization technique in \cite{Benini, Doroud}. In this case the CY manifold $Y$ coincides with the space of the supersymmetric vacua states of this GLSM.
 The JKLMR conjecture was proven some time ago in \cite{Gomis1, Gomis2, Gomis3}.
\par Since  the JKLMR conjecture is right, then due to the mirror symmetry, its mirror version should be right \cite{Bonelli, JKLMR, JKLMRfermat} as well
\begin{equation}
\label{eq:conjmirror}
    Z_Y=e^{-K_C^X(\psi_l)}.
\end{equation} 
Here  $K_C^X(\psi_l)$ is the Kähler potential on $M_C(X)$ - moduli space of complex structures of Calabi-Yau family $X$, which is the mirror partner to $Y$, and  ${\psi_l}$ are  the  complex structure moduli, the coordinates on  $M_C(X)$.
\par Kähler potential $K_C^X(\psi_l)$ for mirror quintic threefold was computed firstly in \cite{CandelasOssa}. Thereafter in \cite{Candelas, Quintic, Fermat, Twomoduli} the special geometry on the moduli space of complex structures $M_C(X)$ has been computed for the wide set of Calabi-Yau manifolds.
\par The mirror version of JKLMR conjecture (\ref{eq:conjmirror}) has been verified \cite{Jockers, JKLMR, JKLMRfermat} for a few cases, see also \cite{Gomis4}.
\par In this work we present a verification of the mirror conjecture (\ref{eq:conjmirror}) for the case that belongs to a  class of Calabi-Yau manifolds  considered by Berglund and Hübsch in \cite{Berglund}. 
\par The key point of our approach is  using the Batyrev's construction \cite{Batyrev}. A Calabi-Yau threefold $X$, defined as a hypersurface in a weighted projective space $\mathbb{P}^4_{(k_1, \dots, k_5)}$ is given by zero locus of quasihomogenious  polynomial $W_X$. Exponents of the monomials in the $W_X$ define finite set $\Vec{V}_a, \ a=1,\dots,N$. Their convex set defines Batyrev's polytope  $\Delta_X$ \cite{Batyrev}. We use the set of vectors $\Vec{V}_a$ for constructing the fan \cite{Mirror}, which defines a toric variety. Calabi-Yau manifold $Y$, which is the mirror of $X$, is realized as a hypersurface in this toric variety by a  homogeneous polynomial $W_Y$. Knowing the fan we find GLSM, its gauge group and the chiral multiplet charges.  The last ones appear as a coefficients of linear relations between the vectors of the fan and define a weights of a toric variety. 

\section{Special geometry for the two-moduli non-Fermat Calabi-Yau}

\qquad In this paper, we consider the  non-Fermat type  manifold $X$ 
 with two deformations  of complex structure \cite{Berglund}.  
The manifold $X$ is defined as a hypersurface in a weighted projective space:
\begin{equation}
  \mathbb{P}^4_{(k_1, \dots, k_5)} = \{(x_1, \dots , x_5) \ | \ (x_1, \dots , x_5)
   \sim (\lambda^{k_1}x_1, \dots ,\lambda^{k_5} x_5) \}.
\end{equation}
\par The main point is that  this model is  not of the Fermat type, which was considered before in \cite{JKLMRfermat}. Namely, let Calabi-Yau $X$ is given by equation in $\mathbb{P}^4_{(3,2,2,7,7)}$ 
\begin{equation}
\label{eq: CY}
    W_{X}(x | \psi)=x_1^7+x_2^7x_4+x_4^3+x_3^7x_5+x_5^3 - \psi_1x_1x_2x_3x_4x_5+\psi_2x_1^3x_2^3x_3^3 =0.
\end{equation}
The degree of the polynomial equals $d=21$. The phase symmetry group of $W_X$ at $\psi_1=\psi_2=0$ is $\mathbb{Z}_{21}^2\times \mathbb{Z}_7$. 

We consider a quotient $\hat{X}=X/H$, where $H :=\left(\mathbb{Z}_{21} : 12,2,0,7,0\right)$. The Hodge numbers of this 2-parameter family are  $h_{1,2}(\hat{X})=2$ and $h_{1,1}(\hat{X})=95$. The basis in the invariant part of Milnor's ring $R$ consists of monomials \cite{Twomoduli}: $R^Q=\langle e_1,\dots,e_6 \rangle$ where $e_1=1, \ e_2=x_1x_2x_3x_4x_5, \ e_3=x_1^3x_2^3x_3^3, \ e_4=e_2^2,\ e_5=e_2e_3$ и $ e_6=e_2^2e_3$.

Special geometry on the moduli space of complex structures is given by Kähler potential, which was computed in \cite{Twomoduli} and can be rewritten in a form

\begin{equation}
\label{eq:kahler}
\begin{aligned}
    e^{-K_C^X(\psi_1,\psi_2)} = \sum_{\Vec{\mu}=(\nu,\mu)} (-1)^{|\Vec{\mu}|}\gamma^3\left(\frac{\nu}{7}\right)\gamma^2\left(\frac{\mu}{7}\right)|\sigma_{\Vec{\mu}}(\psi_1,\psi_2)|^2, 
        \end{aligned}
\end{equation}
here $|\Vec{\mu}|= 3(\nu-1)+2(\mu-1)$
\begin{equation}
\label{eq:Periods}
\begin{aligned}
    \sigma_{\Vec{\mu}}(\psi_1,\psi_2) = \sum_{m,n\in \Sigma_{\Vec{\mu}}}(-1)^m\frac{\Gamma^3\left(\frac{1+n+3m}{7} \right)\Gamma^2\left(\frac{2+2n-m}{7} \right)}{\Gamma^3\left(\frac{\nu}{7} \right)\Gamma^2\left(\frac{\mu}{7} \right)}\frac{\psi_1^n\psi_2^m}{n!m!},
      \end{aligned},
\end{equation}

    \begin{equation}
\begin{aligned}
    \Sigma_{\Vec{\mu}}=\{n,m \in \mathbb{Z}_{\geqslant 0} \ |\ 1+n+3m=\nu \ (\text{mod}) \ 7; \  2+2n-m=\mu \ (\text{mod}) \ 7 \},
    \end{aligned}
\end{equation}

\begin{equation}
    \Vec{\mu}=(\nu,\mu)=(1,2), (2,4), (3,6), (4,1), (5,3), (6,5).
\end{equation}

\section{Gauged Linear Sigma Model}
\qquad For the first time, this model was considered by Witten in \cite{Witten}. It was shown in \cite{Benini, Doroud} that  this model can be defined on $S^2$ with preserving $\mathcal{N}=(2,2)$ supersymmetry. It  allows to compute the exact partition function by supersymmetric localization technique \cite{Benini, Doroud}. 
\par Lagrangian of this model \cite{Witten} is a sum of Yang-Mills Lagrangian $\mathcal{L}_{YM}$, kinetic term $\mathcal{L}_{\text{mat}}$ for matter chiral superfields $\Phi_a, \ a=1,\cdots,N$, Fayet–Iliopoulos term $\mathcal{L}_{FI}$ and the term   with superpotential $W_Y(\Phi_a)$. 
\par We consider the theory with gauge group $G=U(1)^h:=\prod_{l=1}^h U(1)_l$ with $h$ gauge vector  superfields $(V_1,\cdots,V_h)$. 
Supersymmetric vacua space of the potential energy for the scalar fields $\phi_a$ of chiral multiplet is a hypersurface $Y$ in a toric variety for suitable values of Fayet–Iliopoulos parameters $r_l$.
\par The toric varieties themselves form a family depending on parameters $r_l$ and theta angles $\theta_l$. Hypersurfaces $Y$ in each of these toric varieties form a family of the Calabi-Yau manifolds, which depend on the coefficients of the polynomial $W_Y$. The last ones are moduli of the complex structure of $Y$. 
The polynomial $W_Y$ is invariant with respect  to coordinate transformations in the toric variety $\phi_a\rightarrow \lambda^{Q_{al}}\phi_a$. The weights $Q_{al}$ are nothing but charges of $U(1)_l$ action.
\par Potential energy for the scalar fields in this theory is given by \cite{Witten}
\begin{equation}
\label{eq:potential}
    U(\phi)=\sum_{l=1}^h\frac{e_l^2}{2}\left(\sum_{a=1}^NQ_{al}|\phi_a|^2-r_l \right)^2+\frac{1}{4}\sum_{a=1}^N\left| \frac{\partial W_Y}{\partial \phi_a}\right|^2,
\end{equation}
Here we denote by $(e_1,\dots,e_h)$ the coupling constants.
\par Supersymmetric vacua of the theory is given by minima of the potential (\ref{eq:potential}) modulo gauge symmetry. For $r_l\textgreater 0 $ it is defined as a manifold
\begin{equation}
    \label{eq:vacua}
    Y=\left\{(\phi_1,\dots,\phi_N)|\sum_{a=1}^N Q_{al}|\phi_a|^2=r_l, l=1,\dots, h, \ \frac{\partial W_Y}{\partial \phi_a}=0 \right\}/U(1)^h.
\end{equation}
\par An equivalent way \cite{Mirror} to define a manifold (\ref{eq:vacua}) is to set  $\frac{\partial W_Y}{\partial \phi_a}=0$ in a toric variety defined as a quotient
\begin{equation}
    \mathbb{C}^N//(\mathbb{C}^*)^h:=(\mathbb{C}^N - Z)/(\mathbb{C}^*)^h,
\end{equation}
where $Z$ is some $(\mathbb{C}^*)^h$ invariant subset. The charges $Q_{al}$ are weights of the torus action $(\mathbb{C}^*)^h$ on $\mathbb{C}^N$, $\phi_a \rightarrow \lambda^{Q_{al}} \phi_a, \ a=1,\cdots,N, \ l=1,\cdots,h$. That definition is needed for constructing the mirror manifold $Y$ to the initial one $X$ according to the Batyrev construction.
\par Partition function of GLSM was computed exactly using the Supersymmetric Localization and given by the formula \cite{Benini, Doroud}:
\begin{multline}
    \label{eq:partition}
      Z_Y=\sum_{m_l \in \mathbb{Z}} \prod_{l=1}^h e^{-i\theta_l m_l} \int_{C_1}\dots \int_{C_h}\prod_{l=1}^h \frac{d\tau_l}{(2\pi i)} e^{4\pi r_l\tau_l} \prod_{a=1}^N\frac{\Gamma\left(q_a/2+\sum_{l=1}^hQ_{al}(\tau_l-\frac{m_l}{2}) \right)}{\Gamma\left(1-q_a/2-\sum_{l=1}^hQ_{al}(\tau_l+\frac{m_l}{2}) \right)}, 
\end{multline}
where the contours $C_i$ go along the imaginary axis. The parameters $q_a$ denote the R-symmetry charges. Note that partition function (\ref{eq:partition}) does not depend on the coupling constants $e_i$ and the specific choice of the superpotential $W_Y$.
\par The conjecture proposed by Jockers et al \cite{Jockers} is that partition function (\ref{eq:partition}) matches with the exponent of the Kähler potential $e^{-K_K^Y}=Z_Y$ of the Kähler moduli space of Calabi-Yau manifold $Y$, defined as a hypersurface in a toric variety. The last one arises as a manifold of the supersymmetric vacua of the GLSM. (\ref{eq:vacua}). This statement has been checked for a few examples of Calabi-Yau manifolds in \cite{Jockers, JKLMR, JKLMRfermat}. The main problem of verification is the complexity of computing the special Kähler geometry $e^{-K_K^Y}$. 
\par Since the construction of the mirror symmetry implies the equality
\begin{equation}
    \label{eq:mirror}
    K_K^Y=K_C^X, 
\end{equation}
then the conjecture (\ref{eq:conj}), as mentioned above, can be checked in mirror form \cite{Bonelli, JKLMR}
 \begin{equation}
 \label{eq:conjecture}
     e^{-K_C^X}=Z_Y.
 \end{equation}
 
\par  We will verify (\ref{eq:conjecture}) for the case of Calabi-Yau  2-parameter family mentioned above, using the previously computed $K_C^X$ in the work \cite{Twomoduli}.

\section{Mirror symmetry}
\qquad Let us discuss the method for constructing Calabi-Yau threefold $Y$ that is a mirror partner  for the $X$, developed in \cite{JKLMR}. Consider Calabi-Yau manifold $X$ defined by zero locus of the quasi-homogeneous polynomial in a weighted projective space $\mathbb{P}^4_{(k_1, \dots, k_5)}$.
\begin{equation}
\label{eq:calabi}
    W_X(x_1,\dots,x_5 | \psi_1,\dots, \psi_h)=\sum_{a=1}^{h+5}C_a\prod_{i=1}^5x_i^{v_{ai}}, \ \ \ \sum_{i=1}^5k_iv_{ai}=d=\sum k_i.
\end{equation}
Here $\psi_i$ are the coordinates on the moduli space of complex structures $M_C(X)$. In fact, the equation (\ref{eq:calabi}) defines the whole family of Calabi-Yau manifolds, corresponding to the points on the moduli space.
\par The set of exponents $v_{ai}$ corresponds to the coordinates of vectors $\vec{V}_a \in \mathbb{R}^5$, that is $v_{ai}=(\vec{V}_a)_i$. These vectors define  a fan which defines a toric manifold that contents the mirror Calabi-Yau manifold $Y$.  More precisely, they are the edges of this fan. Using this fact we will build the mirror manifold $Y$.\\ These  vectors $\vec{V}_a$ being vectors in five-dimensional space satisfy the linear relations
\begin{equation}
    \sum_{a=1}^{5+h} Q_{al}\vec{V}_a=0 , \ \ \ l=1,\dots, h,
\end{equation}
where $Q_{al}$ is a set of integer numbers that we choose such that they form an integral basis of the linear relations between the exponents of the monomials of $W_X$.
\par So now, using the data $Q_{al}$ we can define a toric variety $\mathbb{C}^N//(\mathbb{C}^*)^h$, $N=h+5$. Namely, consider the coordinates $(\phi_1,\dots, \phi_N)$ in  $\mathbb{C}^N$ and define the factorization
\begin{equation}
    \label{eq:factorization}
    (\phi_1,\dots, \phi_N) \sim (\lambda^{Q_{1l}}\phi_1,\dots, \lambda^{Q_{Nl}}\phi_N), \ \ \ l=1,\dots, h.
\end{equation}
Then the mirror Calabi-Yau manifold $Y$ for $X$ is realized as a hypersurface in the toric variety given by the homogeneous polynomial $W_Y$ such that
\begin{equation}
    W_Y (\lambda^{Q_{1l}}\phi_1,\dots, \lambda^{Q_{Nl}}\phi_N)=W_Y(\phi_1,\dots,\phi_N).
\end{equation}

\section{Verification of the conjecture}
\qquad Let us proceed to the verification of the mirror version of the JKLMR conjecture \cite{Jockers}. Namely, we will show by explicit computation the equality $Z_Y=e^{-K_C^X}$. 
\par We write the polynomial (\ref{eq: CY}) in a form 
\begin{equation}
    W_{X}(x | \psi)=\sum_{a=1}^7C_a\prod_{i=1}^5x_i^{v_{ai}}.
\end{equation}
The  exponents $v_{ai}$ are  coordinates of the vectors $\vec{V}_a \in \mathbb{R}^5$,  that is $v_{ai}=(\vec{V}_a)_i$. Where vectors $\vec{V}_a$ are
\begin{equation}
    \vec{V}_1=
    \begin{pmatrix}
    7 \\
    0 \\
    0 \\
    0 \\
    0
    \end{pmatrix}, 
    \vec{V}_2=
    \begin{pmatrix}
    0 \\
    7 \\
    0 \\
    1 \\
    0
    \end{pmatrix},
    \vec{V}_3=
    \begin{pmatrix}
    0 \\
    0 \\
    0 \\
    3 \\
    0
    \end{pmatrix},
    \vec{V}_4=
    \begin{pmatrix}
    0 \\
    0 \\
    7 \\
    0 \\
    1
    \end{pmatrix},
    \vec{V}_5=
    \begin{pmatrix}
    0 \\
    0 \\
    0 \\
    0 \\
    3
    \end{pmatrix},
    \vec{V}_6=
    \begin{pmatrix}
    1 \\
    1 \\
    1 \\
    1 \\
    1
    \end{pmatrix},
    \vec{V}_7=
    \begin{pmatrix}
    3 \\
    3 \\
    3 \\
    0 \\
    0
    \end{pmatrix}.
\end{equation}
These seven vectors being five-dimensional satisfy two linear relations:
\begin{equation}
\label{eq:linear}
    \sum_{a=1}^7 Q_{al}\vec{V}_a=0 , \ \ \ l=1,2,
\end{equation}
here the $Q_{al}$ is a set of integer numbers such that (\ref{eq:linear}) defines an integer basis in a space of linear relations between vectors $\Vec{V}_a$.
\par The convenient choice of  the $Q_{al}$ is:
\begin{equation}
\label{eq:charge}
    Q_{a1}=(1,1,0,1,0,-1,-2), \ Q_{a2}=(0,0,1,0,1,-3,1).
\end{equation}
\par Let us construct the connection between the model with manifold $X$ and Gauged Linear Sigma Model. Following the approach developed in \cite{JKLMR} we set $h=h_{2,1}=2$, i.e. consider a theory with a gauge group  $U(1)^2$ and $N=5+h=7$ chiral multiplets with charges $Q_{al}$ from the (\ref{eq:charge}).
\par The exact partition function of this model is given by the expression:
\begin{equation}
\begin{aligned}
       Z_Y=\sum_{m_l \in \mathbb{Z}} e^{-i\theta_1 m_1} e^{-i\theta_2 m_2}\int_{C_1}\int_{C_2} \frac{d\tau_1}{(2\pi i)}\frac{d\tau_2}{(2\pi i)} e^{4\pi r_1\tau_1}e^{4\pi r_2\tau_2}\times \\ \times \prod_{a=1}^7\frac{\Gamma\left(q_a/2+Q_{a1}(\tau_1-\frac{m_1}{2})+Q_{a2}(\tau_2-\frac{m_2}{2}) \right)}{\Gamma\left(1-q_a/2-Q_{a1}(\tau_1+\frac{m_1}{2})-Q_{a2}(\tau_2+\frac{m_2}{2})\right)}. 
\end{aligned}       
\end{equation}
\par We set the charges of R-symmetry $q_1=q_2=q_4=2/7, \ q_3=q_5=4/7, \ q_6=q_7=0$. We introduce a change of coordinates on a vacua space
\begin{equation}
\label{eq:map}
\begin{aligned}
    z_1&=e^{-\frac{2\pi}{7}\left[(r_1+2i\theta_1)+2(r_2+2i\theta_2)\right]},
    \\
     z_2&=e^{-\frac{2\pi}{7}\left[3(r_1+2i\theta_1)-(r_2+2i\theta_2)\right]}.
    \end{aligned}
\end{equation}
Then we obtain
\begin{multline}
\label{eq: part}
    Z_Y=\sum_{m_1} \sum_{m_2} \int_{C_1} \int_{C_2}\frac{d\tau_1}{(2\pi i)}\frac{d\tau_2}{(2\pi i)}z_1^{-(\tau_1-\frac{m_1}{2})-3(\tau_2-\frac{m_2}{2})}\bar{z}_1^{-(\tau_1+\frac{m_1}{2})-3(\tau_2+\frac{m_2}{2})}\times \\  \times z_2^{-2(\tau_1-\frac{m_1}{2})+(\tau_2-\frac{m_2}{2})}\bar{z}_2^{-2(\tau_1+\frac{m_1}{2})+(\tau_2+\frac{m_2}{2})}\times \\ \times \frac{\Gamma^3\left(1/7 +(\tau_1-\frac{m_1}{2})\right)\Gamma^2\left(2/7 +(\tau_2-\frac{m_2}{2})\right)}{\Gamma^3\left(1-1/7-(\tau_1+\frac{m_1}{2})\right)\Gamma^2\left(1-2/7-(\tau_2+\frac{m_2}{2})\right)}\times \\ \times 
    \frac{\Gamma\left(-(\tau_1-\frac{m_1}{2})-3(\tau_2-\frac{m_2}{2})\right)}{\Gamma\left(1+(\tau_1+\frac{m_1}{2})+3(\tau_2+\frac{m_2}{2}) \right)}\frac{\Gamma\left(-2(\tau_1-\frac{m_1}{2})+(\tau_2-\frac{m_2}{2})\right)}{\Gamma\left(1+2(\tau_1+\frac{m_1}{2})-(\tau_2+\frac{m_2}{2})\right)}.
\end{multline}
\par For $|z_l|\textgreater 1$ the contours can be deformed to the right half-plane. Then, by  Cauchy theorem, the integrals (\ref{eq: part}) are reduced to the sum of residues at the poles of the gamma function
\begin{equation}
    \left(\tau_1-\frac{m_1}{2}\right)+3\left(\tau_2-\frac{m_2}{2}\right)=n, \ \ 2\left(\tau_1-\frac{m_1}{2}\right)-\left(\tau_2-\frac{m_2}{2}\right)=m , \ \ m,n=0,1,\dots 
\end{equation}
Also denote
\begin{equation}
    \bar{n}=\left(\tau_1+\frac{m_1}{2}\right)+3\left(\tau_2+\frac{m_2}{2}\right), \ \ \bar{m}=2\left(\tau_1+\frac{m_1}{2}\right)-\left(\tau_2+\frac{m_2}{2}\right),
\end{equation}

\begin{equation}
\begin{aligned}
    \left(\tau_1-\frac{m_1}{2}\right)=\frac{n+3m}{7}, \ \ \   \left(\tau_2-\frac{m_2}{2}\right)= \frac{2n-m}{7}, \\ 
     \left(\tau_1+\frac{m_1}{2}\right)=\frac{\bar{n}+3\bar{m}}{7},\ \ \ \left(\tau_2+\frac{m_2}{2}\right)= \frac{2\bar{n}-\bar{m}}{7}. \end{aligned}
\end{equation}
\par When $1+n+3m=0 \ (\text{mod}) \ 7 $ and $2+2n-m=0 \ (\text{mod}) \ 7$ the gamma functions in the denominator of the formula (\ref{eq: part}) have poles,   therefore the corresponding terms in the sum vanish. It follows that the sum in(\ref{eq: part}) effectively goes over the set:
\begin{multline}
    \Sigma_{\Vec{\mu}}=\{n,m \in \mathbb{Z}_{\geqslant 0} \ |\ 1+ n+3m=\nu \ (\text{mod}) \ 7; \ 2+ 2n-m=\mu \ (\text{mod}) \ 7  \},
\end{multline}
\begin{equation}
    \Vec{\mu}=(\nu,\mu)=(1,2), (2,4), (3,6), (4,1), (5,3), (6,5).
\end{equation}

From the relations $\bar{n}+3\bar{m}=n+3m+7m_1, \ 2\bar{n}-\bar{m}=2n-m+7m_2$ we conclude that the numbers $\bar{n}$ and $\bar{m}$ belong to the same classes  $\Sigma_{\Vec{\mu}}$.
\par Therefore the partition function can be rewritten as
\begin{multline}
    Z_Y=\sum_{\Vec{\mu}=(\nu,\mu)} \sum_{(n,m), (\bar{n},\bar{m}) \in \Sigma_{\Vec{\mu}}}z_1^{-n}\bar{z}_1^{-\bar{n}}z_2^{-m}\bar{z}_2^{-\bar{m}}\frac{\Gamma^3\left(\frac{1+n+3m}{7} \right)\Gamma^2\left(\frac{2+2n-m}{7} \right)}{\Gamma^3\left(1-\frac{1+\bar{n}+3\bar{m}}{7} \right)\Gamma^2\left(1-\frac{2+2\bar{n}-\bar{m}}{7} \right)} \times \\ \times \frac{(-1)^n(-1)^m}{n!m!\Gamma\left(1+\bar{n}\right)\Gamma\left(1+\bar{m}\right)}= \\ =\pi^{-5}
    \sum_{\Vec{\mu}=(\nu,\mu)} \sum_{(n,m), (\bar{n},\bar{m}) \in \Sigma_{\bar{\mu}}} (-1)^n(-1)^m\frac{z_1^{-n}\bar{z}_1^{-\bar{n}}z_2^{-m}\bar{z}_2^{-\bar{m}}}{n!m!\bar{n}!\bar{m}!}\times \\ \times \sin^3\left(\pi\frac{1+\bar{n}+3\bar{m}}{7}\right)\sin^2\left(\pi\frac{2+2\bar{n}-\bar{m}}{7}\right)\times \\ \times \Gamma^3\left(\frac{1+n+3m}{7} \right)\Gamma^2\left(\frac{2+2n-m}{7} \right)
    \Gamma^3\left(\frac{1+\bar{n}+3\bar{m}}{7} \right)\Gamma^2\left(\frac{2+2\bar{n}-\bar{m}}{7} \right).
\end{multline}
\par Using the identities: 
\begin{multline}
   \pi^{-5} \sin^3\left(\pi\frac{1+\bar{n}+3\bar{m}}{7}\right)\sin^2\left(\pi\frac{2+2\bar{n}-\bar{m}}{7}\right)=\\ =\pi^{-5} (-1)^{\bar{n}+\bar{m}}(-1)^{|\Vec{\mu}|} \sin^3\left(\frac{\pi \nu}{7}\right)\sin^2\left(\frac{\pi \mu}{7}\right) = \\ =(-1)^{\bar{n}+\bar{m}}(-1)^{|\Vec{\mu}|} \frac{1}{\Gamma^3\left(\frac{\nu}{7}\right)\Gamma^3 \left(1-\frac{\nu}{7}\right)}\frac{1}{\Gamma^2\left(\frac{\mu}{7}\right)\Gamma^2 \left(1-\frac{\mu}{7}\right)}= \\ =(-1)^{\bar{n}+\bar{m}}(-1)^{|\Vec{\mu}|}
   \gamma^3\left(\frac{\nu}{7}\right) \gamma^2\left(\frac{\mu}{7}\right) \frac{1}{\Gamma^6\left(\frac{\nu}{7}\right)}
   \frac{1}{\Gamma^4\left(\frac{\mu}{7}\right)},
\end{multline}
\par we find:
\begin{equation}
    Z_Y=\sum_{\Vec{\mu}=(\nu,\mu)} (-1)^{|\Vec{\mu}|}\gamma^3\left(\frac{\nu}{7}\right)\gamma^2\left(\frac{\mu}{7}\right)|\sigma_{\Vec{\mu}}(z_1,z_2)|^2, 
\end{equation}

\begin{equation}
    \sigma_{\Vec{\mu}}(z_1,z_2) = \sum_{m,n\in \Sigma_{\Vec{\mu}}}(-1)^m(-1)^n\frac{\Gamma^3\left(\frac{1+n+3m}{7} \right)\Gamma^2\left(\frac{2+2n-m}{7} \right)}{\Gamma^3\left(\frac{\nu}{7} \right)\Gamma^2\left(\frac{\mu}{7} \right)}\frac{z_1^{-n} z_2^{-m}}{n!m!}.
\end{equation}
\par The expression for the partition function matches with the formula (\ref{eq:kahler}) for  $e^{-K_C^X}$ from the paper \cite{Twomoduli}. In order to obtain this equality we must identify moduli of complex structures of the Calabi-Yau $X$ $\psi_1, \psi_2$ with Kähler moduli of the manifold $Y$ $z_1, z_2$ as
\begin{equation}
    \psi_1=-z_1^{-1}, \ \ \psi_2=z_2^{-1},
\end{equation}
where $z_1,z_2$ are connected with the parameters  $r_l,\theta_l$ by the formula (\ref{eq:map}).
\par These relations give  the mirror map for the considered case.
\section*{Conclusion}
\qquad Starting from the model with non-Fermat Calabi-Yau $X$ we have constructed the  $\mathcal{N}=(2,2)$ Gauged Linear Sigma Model with the manifold of supersymmetric vacua $Y$, which is the mirror for $X$. Knowing the Special geometry on the moduli space of complex structures on $X$ and using Mirror symmetry in Batyrev's approach \cite{Batyrev} we have checked JKLMR conjecture \cite{Jockers} for this case having obtained the explicit equality $Z_Y=e^{-K_C^X}$. 
We done that for the case of Calabi-Yau $X$ of non-Fermat type which was not considered before in this way. 
\par The formula (\ref{eq: part}) is also important because it gives an analytic continuation for the Kähler potential $K_C^X$ on the moduli space of complex structures outside the region of convergence of the series (\ref{eq:Periods}).
\section*{Acknowledgments}

\par We are grateful to K. Aleshkin, G. Koshevoi, F, Malikov, A. Litvinov, V. Pestun, and M. Kontsevich for the useful discussions.  A. B. is grateful to prof. Pestun for the possibility to visit IHES in 2019.
The project has received funding from the European Research Council (ERC) under the European Union's Horizon 2020 research and innovation program (QUASIFT grant agreement 677368).
This work was done in Landau Institute for Theoretical Physics and has been supported by the Russian Science Foundation under the grant 18-12-00439.


\end{document}